\documentclass[aps,10pt,prd,twocolumn,preprintnumbers,amsmath,amssymb,nofootinbib,superscriptaddress,a4paper,showpacs,tightenlines,longbibliography]{revtex4-2}
\usepackage{balance}
\pdfoutput=1
\newcommand{\eq}[1]{\vspace{-5.5pt}\begin{equation}\hspace{2pt}#1\hspace{-0pt}\vspace{-3.5pt}\end{equation}}
\usepackage{xcolor,titlesec,tikz-cd,datetime2,graphicx,amsfonts,amsthm,bm,bbm,mathrsfs,amsmath,float,pgf,tikz,mathrsfs,braket}
\titleformat{\section}{\centering\normalsize\normalfont\bf}{\thesection}{0em}{}
\usetikzlibrary{cd}
\definecolor{lapis}{rgb}{0.0.0470,0.2941,0.5568}
\definecolor{burgundy}{rgb}{0.5, 0.0, 0.13}
\newcommand{\Y}[3]{{}_{#1}Y_{#2 #3}}
\usepackage[T1]{fontenc}
\usepackage[english]{babel}
\usepackage[utf8]{inputenc}
\usetikzlibrary{arrows}

\newcommand{\mref}[1]{(\ref{#1})}
\newcommand{\ceth}{\overline{\eth}}
\newcommand{\z}{\overline{z}}

\usepackage[colorlinks=true, linkcolor=burgundy, urlcolor=lapis, citecolor=lapis, anchorcolor=green]{hyperref}  

\begin{document}
\title{{\textsc{Spin Multipole Expansion of the Memory Effect}}}
\author{\vspace{-0.25cm}\textsc{Nikhil Kalyanapuram}}
\email{nkalyanapuram@psu.edu}
\affiliation{Department of Physics and Institute for Gravitation and the Cosmos, The Pennsylvania State University, University Park PA 16802, USA}


\begin{abstract}
Using two-dimensional duals of soft dynamics in QED and gravity, we perform an expansion of the respective memory effects in spin multipoles.
\end{abstract}


\maketitle


\vspace{-15pt}
\section*{\textsc{Introduction}}
\vspace{-10pt}
The behaviour of QED and gravity in the deep infrared have long been understood to exhibit rich dynamical characteristics \cite{PhysRev.52.54,Schwinger:1949ra,Yennie:1961ad,Kinoshita:1962ur,Lee:1964is,Jauch:1976ava,Weinberg:1964ew,Weinberg:1965nx,DeWitt:1967uc,Kibble:1968sfb,Kibble:1969ip,Kibble:1969ep,Kibble:1969kd,Kulish:1970ut,Frohlich:1978bf,Frohlich:1979uu,Naculich:2011ry,Akhoury:2011kq,Balachandran:2013wsa,Balachandran:2014hra,R:2018fup}. More recently, it has emerged that the concomitant divergences are closely related to asymptotic symmetries intrinsic to field theory at null infinity on asymptotically-flat spacetimes \cite{Newman:1968uj,Strominger:2013jfa,He:2014laa,He:2014cra,Strominger:2014pwa,Kapec:2014zla,Campiglia:2014yka,Campiglia:2015qka,Campiglia:2015kxa,Campiglia:2016hvg,Campiglia:2016efb,Campiglia:2016jdj,Strominger:2017zoo}. In particular, the soft photon theorem turns out to be closely related to the group of large gauge transformations (read angle-dependent) on the celestial sphere.

The relationship between the nature of QED and gravity deep in the infared and the storage of information at null infinity is captured in an especially clean fashion by the so-called memory effect. The memeory effect records the `change' in the long-range electromagnetic or gravitational fields created by particles participating in a scattering process. Originally derived classically \cite{Braginsky:1985vlg,PhysRevD.44.R2945,PhysRevD.45.520,Christodoulou:1991cr,Blanchet:1992br,Favata:2008yd,Favata:2010zu,Tolish:2014bka}, it has since been realized that it is closely related to the soft theorems in QED and gravity. In particular, at lowest order, the soft theorem computes precisely the memory effect. Indeed, the change in the electromagnetic field or metric captured by the memory effect is given by the corresponding soft theorem. 

The fact that dynamically nontrivial quantities computed by bulk dynamics in QED and gravity are often equivalently defined in terms of intrinsically two-dimensional quantities on the celestial sphere has led to a larger goal of finding dual holographic descriptions of these phenomena, in the hope of eventually realizing a natural example of flat-space holography. Recently, work by the present author \cite{Kalyanapuram:2020epb,Kalyanapuram:2021bvf,Kalyanapuram:2021tnl} has suggested that at least in the infrared, the dynamics of QED and gravity are naturally encoded in a class of two-dimensional models, so long as one restricts attention to massless particles when dealing with scattering. At leading order in particular, QED and gravity in the infrared are well-described by making use of the free boson and biharmonic scalar in two dimensions respectively. 

In the language of these two-dimensional theories, the soft theorems are equivalent to a class of Ward identities corresponding to the global symmetries of the models. Accordingly, it becomes possible to recast the memory effect as a statement of operator product expansions; the memory effect is equivalent to inserting the Noether currents of these global symmetries. In this work, we will leverage this interpretation of the memory effect to perform a multipole expansion of electromagnetic and gravitational memories. Due to the intrinsically two-dimensional nature of the models we use, it becomes relatively straightforward to make full use of the technology of spin-weighted spherical harmonics \cite{Janis:1965tx,Lamb1966,Newman:1966ub,Goldberg:1966uu}, allowing us to write down simple expressions for the memories in terms of these functions. 

\vspace{-15pt}
\section*{\textsc{Soft Theorems from Two-Dimensional Ward Identities}}
\vspace{-10pt}
The key observation we find most helpful in expanding the soft theorems in multipoles in the realization of soft theorems at leading order in QED and gravity as Ward identities of two-dimensional models. Accordingly, it is best to start with the derivation of this duality to inform the corresponding expansion. 

The two-dimensional dual models for QED and gravity take the form of free-boson theories on the celestial sphere. For QED we have the standard free boson theory in two-dimensions---

\eq{\label{eq:1}S^{(1)} = -\int dz\wedge d\z\left(\eth\varphi(z,\z)\ceth\varphi(z,\z)\right)}
where $\varphi(z,\z)$ is the dual photon on the sphere. To keep everything completely covariant, we have chosen to express the action in terms of the covariant derivatives on the sphere; the specialization to the plane is readily made by replacing all derivatives with their flat counterparts.

The soft $S$-matrix is encoded in two-dimensional correlators dual to scattering amplitudes by dressing external single particle operators (denoted by $\mathcal{O}(\omega_{i},z_{i},\z_{i})$) via the prescription

\eq{\mathcal{O}(\omega_{i},z_{i},\z_{i}) \longrightarrow \mathcal{W}^{(1)}(e_{i},z_{i},\z_{i})\mathcal{O}(\omega_{i},z_{i},\z_{i})}
where

\eq{\mathcal{W}^{(1)}(e_{i},z_{i},\z_{i}) = \exp(ie_{i}\varphi(z_{i},\z_{i})).}
The soft theorem then boils down to the simple statement we have a Ward identity, namelt that due to the shift symmetry of the action in \mref{eq:1} enjoys a symmetry under translations according to $\varphi\longrightarrow\varphi+a$. In other words, we have Noether currents defined by

\eq{j^{(1)}(z) = \eth\varphi(z,\z)}
and its conjugate. From the operator product expansion

\eq{\eth\varphi(z,\z)\varphi(z',\z') = \frac{1}{\pi}(1+z\z)\frac{1}{z-z'}}
we infer the Ward identity

\eq{\begin{aligned}
&j^{(1)}(z)\prod_{i}\mathcal{W}^{(1)}(e_{i},z_{i},\z_{i})\sim\\ &\frac{i}{\pi}(1+z\z)\frac{e_{i}}{z-z_{i}}\prod_{i}\mathcal{W}^{(1)}(e_{i},z_{i},\z_{i}).
\end{aligned}}
We see that the Ward identity corresponding to shift symmetries imply as direct consequences the soft theorems for the radiation of positive or negative helicity photons (the negative helicity case is of course obtained by the replacement $\eth\rightarrow\ceth$). As an upshot, we see that the shift of the field $\varphi$ by a constant leads to the introduction of an overall phase $\delta_{1}$, given by the sum

\eq{\delta_{1} = e_{1}+\dots +e_{n}}
which is simply the sum of the charges of the external states. Indeed, to ensure single-valuedness of the correlation function of the exponential operators, this phase must vanish, which is equivalent to the statement that charge is globally conserved.

Repeating this analysis for gravity is instructive. The action of the soft modes is entirely analogous to that of QED, except that the derivatives are of order two, rather than one---

\eq{S^{2} = \int dz\wedge d\z\left(\eth^{2}\sigma(z,\z)\ceth^{2}\sigma(z,\z)\right).}
Operators creating external states are once again dressed by exponential operators. Where the QED operators depended on the charges of the external states, the gravitational dressing functions now take into consideration the energies of the scattering particles via

\eq{\label{eq:9}\mathcal{W}^{(2)}(\omega_{i},z_{i},\z_{i}) = \exp\left(i\kappa\omega_{i}\sigma(z_{i},\z_{i})\right).}
The global symmetries of the two-dimensional model of soft graviton are inferred by looking at shifts $f$ that obey the equations

\eq{\eth^{2}f = \ceth^{2}f = 0}
which are satisfied by four-parameter functions given by

\eq{f(z,\z) = a_{1}+a_{2}z + a_{3}\z + a_{4}z\z.}
Indeed, one can guess the conservation laws implied by such symmetries. A product of $n$ Wilson lines as given by \mref{eq:9} will receive a correction $\exp(i\delta_{2})$ when the field $\sigma$ is shifted according to the preceding four-parameter group. This phase will simply be

\eq{\delta_{2} = \kappa\sum_{i=1}^{n}\left(\omega_{i}a_{1}+\omega_{i}z_{i}a_{2}+\omega_{i}\z_{i}a_{3}+\omega_{i}z_{i}\z_{i}a_{4}\right)}
which we need to have vanish if we are to ensure single-valuedness. Since each coefficient of the $a_{i}$`s must vanish independently, we find that the vanishing of the preceding phase necessarily equates to the conservation of global four-momentum.

It turns out that just like in the case of QED, the Noether currents in the gravitational analogue naturally lead to the soft graviton theorems at leading order. Indeed, we have the Noether currents 

\eq{j^{(2)}(z,\z) = \eth^{2}\sigma(z,\z)}
and the conjugate obtained from the replacement $\eth\rightarrow\ceth$. This time, we make use of the operator product expansion

\eq{\eth^{2}\sigma(z,\z)\sigma(z',\z') = \frac{1}{\pi}\frac{1+z\z}{1+z'\z'}\frac{\z-\z'}{z-z'}}
(which can be seen from the fact that the right side is really just the Green's function of $\ceth^{2}$ \cite{Campiglia:2015kxa,H:2018ktv}; it has spin-weight 2). Indeed, this is just the soft factor for the radiation of a single positive helicity soft graviton when the external particle has direction $(z',\z')$ on the celestial sphere.

These currents can now be inserted into a correlation function of the exponential dressing operator due to soft gravitons; we have for example in the case of the positive helicity insertion the identity

\eq{\begin{aligned}
&j^{(2)}(z)\prod_{i}\mathcal{W}^{(2)}(\omega_{i},z_{i},\z_{i})\sim\\ &\sum_{i=1}^{n}\frac{i}{\pi}\omega_{i}\frac{1+z\z}{1+z_{i}\z_{i}}\frac{\z-\z_{i}}{z-z_{i}}\prod_{i}\mathcal{W}^{(2)}(\omega_{i},z_{i},\z_{i})
\end{aligned}}
which can be recognized as the soft theorem due to the emission of a positive helicity graviton.

The two-dimensional representation of the soft theorems together with the covariant form of presentation make it possible for us to perform an expansion using as basis functions the spherical harmonics. Doing so will tell us what the multipole structure of the soft theorems would look like. Performing this expansion will be the task we take up now.

\vspace{-15pt}
\section*{\textsc{Expanding the Soft Theorems in Spin Multipoles}}
\vspace{-10pt}
In order to consistently expand the soft theorems in basis functions, the most important first task at hand is determining the right set of basis functions. Since we are ultimately working on the sphere, it is natural to assume that we will have to use the well-known (and well-understood) spherical harmonics, or a more convenient cousin thereof. 

To figure out the correct expansion we have to use, it is best to look at the spin weights of the current we want to expand. Let us take a look first at the positive helicity current in QED. Note that the positive helicity current in QED is given by the operator $\eth\varphi(z,\z)$. Now the field $\varphi$ is to be regarded as a free scalar, and correspondingly is to have spin weight zero. Indeed, what this means is that the insertion of one instance of the current $j^{(1)}$ amounts to the insertion of an operator of spin-weight one, as the operator $\eth$ is a raising operator in the spin basis.

To perform an expansion of such a function, we make use of the so-called spin-weighted spherical harmonics, defined according to the recursion relations given by

\eq{\eth\Y{s}{\ell}{m} = \sqrt{(\ell-s)(s+\ell+1)}\Y{s+1}{\ell}{m},}
and
\eq{\ceth\Y{s}{\ell}{m} = -\sqrt{(s+\ell)(\ell-s+1)}\Y{s-1}{\ell}{m}.}
To recall, the spin operators $\eth$ and $\ceth$ can be thought of as covariant derivatives projected onto the null vectors $m^{\mu}$ and $\overline{m}^{\mu}$ used to expand the complex metric on the sphere. At the level of the spherical harmonics, they serve to raise or lower the net spin weight. Importantly, it is generally not possible to perform integration by parts with respect to these operators. However, when the total spin weight under the integral sign vanishes, integration by parts can be carried out with no problems. Accordingly, keeping this simple point in mind to start with, we see that the right choice of basis function for the positive helicity current would be $\Y{1}{\ell}{m}$, or equivalently, we may define modes according to

\eq{j^{(1)}_{\ell m} = \int \Y{1}{\ell}{m}^{*}(\Omega)j^{(1)}(\Omega)d\Omega}
where we have schematically expressed the integral in spherical, as opposed to stereographic, coordinates. In the absence of punctures at infinity, we can now comfortably integrate by parts, converting the latter definition into one that is more tractable---

\eq{j^{(1)}_{\ell m} = -\frac{1}{\sqrt{\ell(\ell+1)}}\int Y_{\ell m}^{*}(\Omega)\ceth\eth\varphi(\Omega)d\Omega.}
Acting this on the product of Wilson lines gives precisely the multipole coefficient for quantum numbers $(\ell,m)$---

\eq{\begin{aligned}
&j^{(1)}_{\ell m}\prod_{i}\mathcal{W}^{(1)}(e_{i},z_{i},\z_{i})\sim\\ &-\sum_{i=1}^{n}\frac{ie_{i}}{\pi\sqrt{\ell(\ell+1)}}\int Y_{\ell m}(\Omega)\eth\ceth\langle{\varphi(\Omega)\varphi(\Omega_{i})\rangle}\\
&\times\prod_{i}\mathcal{W}^{(1)}(e_{i},z_{i},\z_{i})
\end{aligned}}
where we have denoted by $\Omega_{i}$ the angular coordinate representation of the stereographic coordinates $(z_{i},\z_{i})$. We now make use of the Green's function representation

\eq{\eth\ceth\langle{\varphi(\Omega)\varphi(\Omega_{i})\rangle} = \pi\delta(\Omega-\Omega_{i})}
to obtain (after a few steps of simplification)

\eq{\begin{aligned}
    &\sum_{i=1}^{n}\frac{ie_{i}}{\pi\sqrt{\ell(\ell+1)}}\int Y_{\ell m}(\Omega)\eth\ceth\langle{\varphi(\Omega)\varphi(\Omega_{i})\rangle} = \\
    &\sum_{i=1}^{n}\frac{ie_{i}}{\sqrt{\ell(\ell+1)}}Y_{\ell m}(\Omega_{i}).
\end{aligned}}
Collecting these facts, we see that the soft factor in QED (denoted by $\mathcal{S}^{(1)}_{\pm}$) is equivalently written (this time in terms of the momentum vectors) as

\eq{\mathcal{S}^{(1)}_{\pm} = \sum_{\ell,m}\frac{\mp1}{\sqrt{\ell(\ell+1)}}\Y{\pm 1}{\ell}{m}(\mathbf{k})M_{\ell m}}
where $M_{\ell m}$ is the multipole moment

\eq{M_{\ell m} = \sum_{i=1}^{n}e_{i}Y^{*}_{\ell m}(\mathbf{k}_{i}).}
It's interesting to see that the coefficients of the basis functions in the spin-weighted spherical basis are just the electric multipole moments of the external particles. More specifically, they are really the changes in the multipole moment when evaluated between the initial and final charge configurations. At the level of these moments, we see that the change in monopole moment is conspicuously absent, given the fact that the spin weighted spherical harmonics for $\ell = 0$ and $s=\pm 1$ vanish identically. The fact that the multipole expansion starts at $\ell=1$ corresponds well to the expectation that one retains from classical intuition, namely that charge monopole moments must always remain conserved. This presents itself here as a statement of charge conservation; the total initial and final electric charges cannot differ from one another.

The generalization to gravity proceeds by identifying the correct spin weight once again, which can be inferred either by analogy or by direct inspection. Indeed, just as the spin weights of the currents in QED were $\pm 1$ due to the action of the operators $\eth$ and $\ceth$ respectively, the weights of the soft currents in gravity will be $\pm2$ due to the quadratic nature of the corresponding derivatives. In accordance with this, we define the operators

\eq{j^{(2)}_{\ell m} = \int \Y{2}{\ell}{m}^{*}(\Omega)j^{(2)}(\Omega)d\Omega}
and its conjugate (under the replacement of $\eth$ with its conjugate and changing the spin weight of the spherical harmonic to $2$). The effect of these operators on the product of Wilson lines can be found by direct evaluation; skipping the more tedious details we find the result for the coefficients of the functions $\Y{\pm 2}{\ell}{m}$

\eq{\begin{aligned}
    &\sum_{i=1}^{n}\frac{i\kappa\omega_{i}}{\pi\sqrt{(\ell-1)\ell(\ell+1)(\ell+2)}}\int Y_{\ell m}(\Omega)\eth^{2}\ceth^{2}\langle{\sigma(\Omega)\sigma(\Omega_{i})\rangle} = \\
    &\sum_{i=1}^{n}\frac{i\kappa\omega_{i}}{\sqrt{(\ell-1)\ell(\ell+1)(\ell+2)}}Y_{\ell m}(\Omega_{i}).
\end{aligned}}
Ultimately, we have for the expansion of the soft factor at leading order in gravity the following multipole expansion in terms of the spin-weighted spherical harmonics---

\eq{\mathcal{S}^{(2)}_{\pm} = \sum_{\ell,m}\frac{1}{\sqrt{(\ell-1)\ell(\ell+1)(\ell+2)}}\Y{\pm 2}{\ell}{m}(\mathbf{k})G_{\ell m}}
where we have again the multipole moment

\eq{G_{\ell m} = \kappa\sum_{i=1}^{n}\omega_{i}Y_{\ell m}(\mathbf{k}_{i})}
which this time measures the difference in the energy multipole for each value of the pair $(\ell,m)$ between the initial and final states in a given scattering process. Once again, the same features of conservation in agreement with classical considerations are manifest; the multipole moments for $\ell<2$ do not contribute, which can be seen in two distinct ways. The simplest of course is that once again, they are trivial for $s=\pm 2$. More fundamentally, the $G_{\ell m}$ valish for $\ell<2$, which is nothing more than a statement of energy conservation ($\ell = 0$) and the conservation of global three-momentum ($\ell=1$). 

We remark finally on how the expansion of the soft theorem in gravity naturally exhibits the relationship between the leading order soft theorem and supertranslations on null infinity. Note that the energies $\omega_{i}$ are conjugate to the locations $u_{i}$ representing retarded time parameters on future null infinity (assuming for the sake of simplicity only outgoing radiative modes). The multipole moments are then operator valued, but in a very specific manner. Indeed, the multipole moments represent precisely the net supertranslational effect due to the presence of hard modes of the coordinates $u_{i}$, with the angle dependent coefficients simply being the corresponding harmonics $Y_{\ell m}(\mathbb{k}_{i})$ at each order in the spherical basis expansion. 
\vspace{-15pt}
\section*{\textsc{Discussion}}
\vspace{-10pt}
In this article, we have presented an expansion of the leading order soft theorems in QED and gravity---known to be equivalent to electromagnetic and gravitational memory respectively---in terms of spin-weighted spherical harmonics by making use of a class of dual models known to capture the infrared dynamics of these theories at leading order. In particular, we have found that the coefficients of the spin-weighted spherical harmonics are given by the differences in multipole moments between the initial and final state. Quite pleasingly, the memory effect seems to naturally encode the change in monopole moment, starting with the dipole moment in QED and with the quadrupole moment in gravity. 

A natural point forward would be the extension of this method of expansion beyond the leading order in the soft expansion. The subleading soft theorems lead to a higher order memory effect, and an expansion in spherical harmonics may be instructive. 

\vspace{-15pt}
\section*{\textsc{Acknowledgements}}
\vspace{-10pt}
I thank Jacob Bourjaily and Alok Laddha for their comments. Partial support for this research is due to ERC Starting Grant (No. 757978) and a grant from the Villum Fonden (No. 15369). This work has been supported in part by the US Department of Energy (No. DE-SC00019066).
\vspace{-10pt}


\bibliographystyle{utphys}
\bibliography{v1.bib}

\end{document}